\documentclass[conference]{IEEEtran}
\IEEEoverridecommandlockouts
\usepackage{cite}
\usepackage{amsmath,amssymb,amsfonts}
\usepackage{algorithmic}
\usepackage{graphicx}
\usepackage{textcomp}
\usepackage{xcolor}

\usepackage{todonotes}
\setuptodonotes{inline}

\usepackage{tabularx}
\usepackage{hyperref}

\usepackage[framemethod=tikz]{mdframed}
\newmdenv[
  innerleftmargin=7pt,
  innerrightmargin=7pt,
  tikzsetting={draw=black,dashed,line width=0.5pt,dash pattern = on 4pt off 2pt},
  linecolor=white,
  backgroundcolor=white
]{dashedbox}
\newmdenv[
  innerleftmargin=7pt,
  innerrightmargin=7pt,
  tikzsetting={draw=black, line width=0.5pt},
  linecolor=black,
  backgroundcolor=white
]{normalbox}
\newmdenv[
  roundcorner=5pt,
  innerleftmargin=7pt,
  innerrightmargin=7pt,
  tikzsetting={draw=black, line width=0.5pt},
  linecolor=black,
  backgroundcolor=white
]{roundedbox}

\newmdenv[
  linewidth=2pt,
  roundcorner=5pt,
  innertopmargin=0pt,
  innerbottommargin=0pt,
]{myframe}

\usepackage[final]{changes}

\def\BibTeX{{\rm B\kern-.05em{\sc i\kern-.025em b}\kern-.08em
    T\kern-.1667em\lower.7ex\hbox{E}\kern-.125emX}}

\begin{document}

\title{Visually Analyzing Company-wide Software Service Dependencies: An Industrial Case Study}

\author{
\IEEEauthorblockN{Sebastian Baltes, Brian Pfitzmann, Thomas Kowark}
\IEEEauthorblockA{SAP SE, Germany\\
\{sebastian.baltes,brian.pfitzmann,thomas.kowark\}@sap.com}
\and
\IEEEauthorblockN{Christoph Treude}
\IEEEauthorblockA{Uni Melbourne, Australia\\
christoph.treude@unimelb.edu.au}
\and
\IEEEauthorblockN{Fabian Beck}
\IEEEauthorblockA{Uni Bamberg, Germany\\
fabian.beck@uni-bamberg.de}
}

\maketitle

\begin{abstract}
Managing dependencies between software services is a crucial task for any company operating cloud applications.
Visualizations can help to understand and maintain these complex dependencies.
In this paper, we present a force-directed service dependency visualization and filtering tool that has been developed and used within SAP.
The tool's use cases include guiding service retirement as well as understanding service deployment landscapes and their relationship to the company's organizational structure.
We report how we built and adapted the tool under strict time constraints to address the requirements of our users.
We further share insights on how we enabled internal adoption.
For us, starting with a minimal viable visualization and then quickly responding to user feedback was essential for convincing users of the tool's value.
The final version of the tool enabled users to visually understand company-wide service consumption, supporting data-driven decision making.
\end{abstract}


\section{Introduction}

Services enable access to capabilities through clearly defined interfaces~\cite{oasis-soa-rm2006}.
The notion of software applications as services gained popularity in the late 1990s with Amazon's service-based model~\cite{Vogels2022} and later as part of the service-oriented architecture paradigm~\cite{oasis-soa-rm2006}.
Since then, different techniques for defining interfaces and consuming services have emerged (e.g., SOAP or REST).
Moreover, the appropriate size of services, ranging from small microservices to large monoliths~\cite{DBLP:conf/icsm/FritzschB0Z19}, has been controversially discussed.

In modern cloud environments, there is often a multitude of services that interact via interfaces, forming complex software systems offered to customers as Software-as-a-Service (SaaS) solutions.
Especially in the context of business software, such solutions are business-critical for customers, requiring a high degree of stability and reliability.
Therefore, managing dependencies between hundreds of services constituting the backbone of SaaS solutions becomes business-critical, too.
For instance, before retiring a previously deprecated service, one must ensure that all dependencies on it have been migrated.
Reliable service metadata is imperative for handling such deprecation scenarios.
A visualization of service dependencies and related metadata can support monitoring a company's service landscape, enabling informed decisions on required maintenance activities.

In this paper, we present insights from an industrial case study at SAP.
We developed a tailored node link visualization that enables the exploration of dependencies and metadata of SAP-managed services across different scenarios.
Our perspective is that of a team working on strategic projects, i.e., executive projects with high priority. The first three authors were part of this team.
We report on an initial version of the visualization that we built for analyzing SAP's cloud service consumption, and a revised version built for investigating service release stages and organizational aspects.
We also discuss factors that supported internal adoption and provide information on how a large software company maintains and analyzes its service metadata and dependencies.
In summary, we report a case study on the productive use of software visualization in industry, which is, according to van Deursen, its ``ultimate measure of success''~\cite{DBLP:conf/softvis/Deursen10}.

\begin{figure}
    \centering
    \includegraphics[width=0.79\linewidth,trim=0 5pt 0 0,clip]{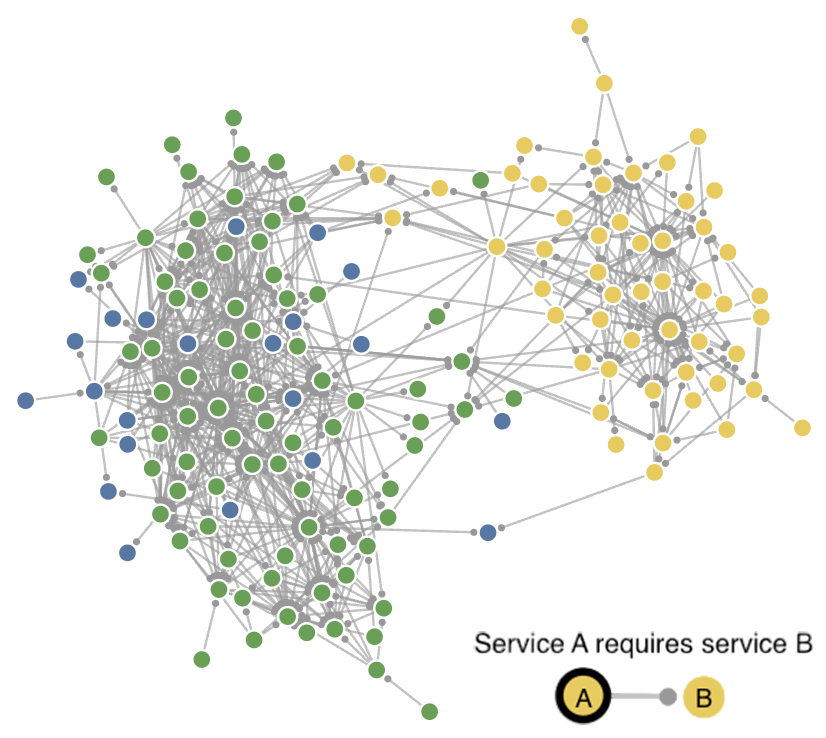}
    \caption{SAP-managed service dependencies of a large organizational unit. Node color encodes native cloud environment of a service. Two large clusters correspond to the older (yellow) and newer (green, blue) environments.}
    \label{fig:all-services}
\end{figure}

\section{Related Work}
\label{sec:related-work}

Research on software dependencies is prevalent, with studies reporting findings such as an exponential increase in inter-project dependencies in contrast to linear project growth~\cite{bavota2013evolution}, risks within software ecosystems due to transitive package dependencies~\cite{kikas2017structure}, and developers' hesitancy to update package dependencies~\cite{kula2018developers}. 
However, such package dependencies are inherently different from service dependencies.
While static dependencies can be derived from package configuration files (e.g., Maven's \texttt{pom.xml}), service dependencies might manifest as a simple HTTP GET request in the source code, accessing a remote service URL.
Therefore, monitoring and controlling complex service dependencies usually involves combining data from multiple sources~\cite{DBLP:journals/queue/EsparrachiariRR18}.

Graph-based visualization of package dependencies is a well-researched topic in software visualization~\cite{10.1016/j.jss.2014.03.071} 
with two main approaches: (i)~node-link diagrams, where dependencies are drawn as links connecting the software artifacts, and (ii)~matrices, which visually indicate adjacency information in the cells of a quadratic matrix~\cite{DBLP:phd/de/Beck2013}. 
As matrix-based approaches can be less accessible and more difficult to apply, node-link approaches---like the one we use---appear to dominate both the research and tool landscape. They can flexibly encode additional metrics~\cite{10.1109/vissof.2011.6069454}, and clearly reveal dense clusters when using a force-directed layout, i.e., a simulated physical system of forces between nodes~\cite{10.1145/1409720.1409735}. 
Although most approaches do not directly address service dependencies, the qualities of force-directed node-link visualizations can be transferred.

Regarding service dependencies, researchers have proposed approaches based on trace analysis to understand the underlying architecture, detect antipatterns, and debug microservice-based systems~\cite{zhou2018fault, DBLP:conf/sigsoft/Guo0WLJDXS20}. 
Researchers have also emphasized the need for anomaly detection in service dependency graphs~\cite{DBLP:conf/bir/GaidelsK20} and proposed graph-based approaches to detect cyclic dependencies~\cite{10.1109/ficloud57274.2022.00042}. 
Specific approaches for visualizing service dependencies have recently been discussed in a survey~\cite{10.1109/sose55356.2022.00011}, where graph-based visualization methods provide the majority of examples.
Although industrial tools exist, most papers are not based on industrial data and focus on individual, often microservice-based, applications.
Our paper contributes a company-wide perspective, which includes understanding and improving dependencies across a diverse set of company-managed services, ranging from small microservices to larger applications and spanning multiple cloud environments.

\section{Background}
\label{sec:background}

This section provides background information on the service metadata that SAP maintains and introduces the use cases that our tool was required to support.

\subsection{Service Metadata at SAP}
\label{sec:service-metadata}

SAP is one of the largest providers of enterprise software both on-premises and in the cloud.
In addition to standalone Software as a Service (SaaS) solutions, SAP provides various other types of services on its cloud platform, the SAP Business Technology Platform (BTP).
Service types include \emph{infrastructure services} (e.g., observability services), \emph{technical services} (e.g., database services), and \emph{application services} (e.g., SAP's SaaS offerings). 
Metadata for all customer-facing and all internal services is collected from different source systems and compiled into a central database that is accessible via an (internal) API and a web tool.
The collected metadata includes a service ID, name, description, the responsible organizational unit, and the service's release stage.
Dependencies on other services and the service's native cloud environment, i.e., the environment the service is deployed to, are also maintained.
Service cross-consumption between cloud environments is possible, but each service has exactly one native cloud environment (e.g., the SAP BTP Cloud Foundry environment).
Service dependencies are modeled as a \emph{requires} relationship, i.e., a service declares its dependencies on other services.
The service release stages model a life cycle from planning over internal and beta usage to general availability (GA). After a service is released, it can be marked as deprecated and later retired.
For our visualization, we (continuously) retrieved metadata for hundreds of services across all cloud environments and release stages via the above-mentioned service metadata API.

\begin{table}
\caption{Main use cases of our visualization tool.}
\label{tab:use-cases}
\begin{tabularx}{\columnwidth}{lX}
\hline
\textbf{Use Case} & \textbf{Description} \\
\hline
\textsc{Consumption} & Enabling a holistic analysis of service deployment and consumption. \\
\textsc{Quality} & Revealing data quality issues. \\
\textsc{Deprecation} & Retiring deprecated services. \\
\textsc{Organization} & Assessing the relationship of service dependencies and the company's organizational structure. \\
\textsc{Reporting} & Preselecting data for reports (filtering using visualization, then additional analyses/reporting in other tools).\\
\hline
\end{tabularx}
\end{table}


\subsection{Use Cases}
\label{sec:use-cases}

Table~\ref{tab:use-cases} lists the main use cases that guided the development of our visualization tool.
The \textsc{Consumption} use case captures the goal of understanding service deployment and consumption.
A cross-cutting concern we were interested in throughout the project was better understanding the \textsc{Quality} of the available service metadata.
Since the service metadata in the above-mentioned database originates from multiple source systems, data quality issues can arise due to different metadata standards, missing information in the source systems, or problems during data processing and aggregation.

After the initial version of our visualization tool was available, colleagues approached us with additional use cases.
While the \textsc{consumption} and data \textsc{quality} aspects were important to them as well, they were particularly interested in better understanding service \textsc{deprecation} and understanding which \textsc{organizations} provide and consume deprecated (but not yet retired) services.
SAP's organization is structured along organizational levels, starting with the top-level board areas down to individual teams.

An important first step towards retirement is marking a service as deprecated, giving dependent services enough time for migration.
If an already deprecated (internal) service is rarely used, it makes sense to retire it soon, enabling the re-prioritization of development and operation resources.
Migration and retirement are easier to orchestrate if services are organizationally close to each other, connecting \textsc{deprecation} and \textsc{organization}.
One of our colleagues' goals was to quickly get a company-wide overview of which services depend on deprecated services, summarizing their findings in a detailed report.
The related \textsc{reporting} use case motivated us to provide interfaces for exporting filtered data from the visualization, enabling users to import that data into analytics tools for additional analysis/visualization (e.g., tables, diagrams).

\section{Visualizing Service Metadata}
\label{sec:visualization}

The existing SAP-internal service metadata tool allowed users to filter services and provided a service dependency visualization.
However, the tool was developed for maintaining and exploring the metadata and dependencies of individual services rather than performing holistic analyses across the entire company.
Teams independently manage their services' dependencies using this central tool, which lends itself to the distributed nature of a service-oriented architecture.
However, there is still a need for global monitoring.
This is where software visualization comes into play.

Although global analyses were possible with existing data and tools, it was more cumbersome than the visual approach we implemented.
For example, our tool made it easier for users to detect service consumption patterns.
To efficiently address the \textsc{Consumption} use case, we quickly developed a first dependency visualization that also allowed assessing data \textsc{Quality}.
To facilitate distribution within the company, the visualization needed to be accessible through a web browser, usable by business users, and support interactive exploration of the data.
Unfortunately, existing graph visualization tools such as \texttt{Gephi} and \texttt{yEd} did not meet these criteria.
Therefore, we decided to implement a custom visualization tool using \texttt{D3.js}.
We developed a first version based on the requirements outlined above in a short period of time, and then later extended the visualization tool to cover more use cases.

\section{Initial Visualization Tool}
\label{sec:visualization-initial}

In the following, we outline central design decisions for addressing the \textsc{Consumption} and \textsc{Quality} use cases.

\subsection{Design Decisions}

Our first fundamental decision was to use a force-directed graph layout so that clusters of connected services as well as completely unconnected services would immediately be visible.
To this end, we utilized \texttt{d3-force} 
for implementing a global many-body force simulation with a negative node strength leading to a node repulsion analog to electrostatic charge.
For a space-efficient representation of services, we further decided to use a circular node shape with the node color encoding one information at a time, opposed to multiple text fields within rectangular nodes in the existing visualization.
Moreover, we needed to draw the directed edges between the nodes in a way that does not clutter the high-level service dependency view too much.
After exploring different arrow forms, we ended up using a small circle to indicate the direction of an edge, which introduced significantly less clutter than arrow heads in dense parts of the graph while still indicating the direction upon close inspection (see Figure~\ref{fig:all-services}). 

\begin{figure}
    \centering
    \includegraphics[width=\linewidth,trim=0 0 0 0,clip]{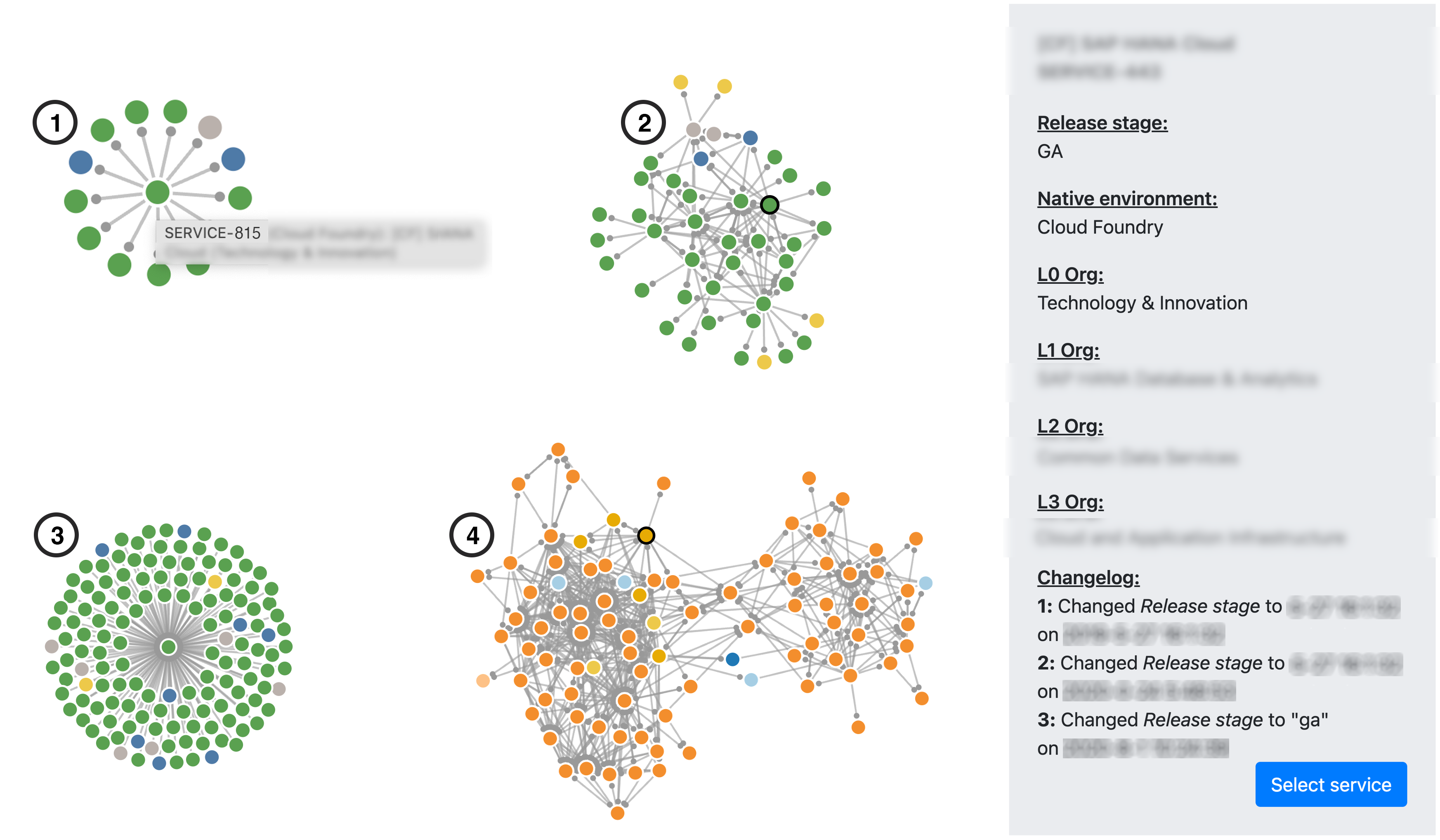}
    \caption{Our visualization tool with different filter and display settings for a selected service: (1) direct dependencies, 
    (2) all dependencies up to a depth of two, (3) inverted dependency relationship, (4) color configured to encode organizational unit instead of native cloud environment.} 
    \label{fig:service-815-viz}
\end{figure}

By default, our visualization only showed the dependencies (as edges) and native cloud environments (encoded in node color).
Additional information was available in a tooltip that appeared while hovering over a service node (see Figure~\ref{fig:service-815-viz}-1), showing a brief summary of the service's metadata (service ID, name, native cloud environment, and the board area responsible for the service).
We further implemented the option to dynamically change the color coding of the nodes to, e.g., project the organizational unit instead of the cloud environment onto the nodes (see Figure~\ref{fig:service-815-viz}-4).
To enable flexible exploration of the data, we added filters that allowed users to focus on the dependencies of a particular service or on all services of a particular organizational unit.
Color coding and filters were stored as URL parameters. 

\subsection{Observations}
\label{sec:initial-observations}

Using our visualization tool, we explored the service landscape and wrote a brief report with our findings on \textsc{Consumption} and data \textsc{Quality}.
One finding was that the two clusters appearing in larger graphs (see, e.g., Figures~\ref{fig:all-services}) were caused by high coupling between services within the two main native cloud environments.
The links between those clusters correspond to service cross-consumption between the cloud environments.
In the following, we refer to cloud environments by their node color.
The yellow cloud environment represents the oldest environment, for which SAP already provides guidance for migrating services to newer environments.
The green cloud environment is more recent, and most services are deployed into it.
The blue environment is the most recent one.

In Figure~\ref{fig:all-services}, which is typical for many organizational units we analyzed, one notices that cross-consumption between services in the recent environments (green and blue) is common and that services in the legacy environment (yellow) form a separate cluster.
Cross-consumption between newer and older services exists (usually green services consuming yellow services) and is caused by yellow services that have not been migrated yet (the environment is still supported and maintained by SAP).
Considering the co-existence of green and yellow, it is understandable that services in the more recent environment (green) consume services in the older one (yellow). 

\begin{figure*}
    \centering
    \includegraphics[width=\linewidth,trim=1pt 5pt 2pt 5pt,clip]{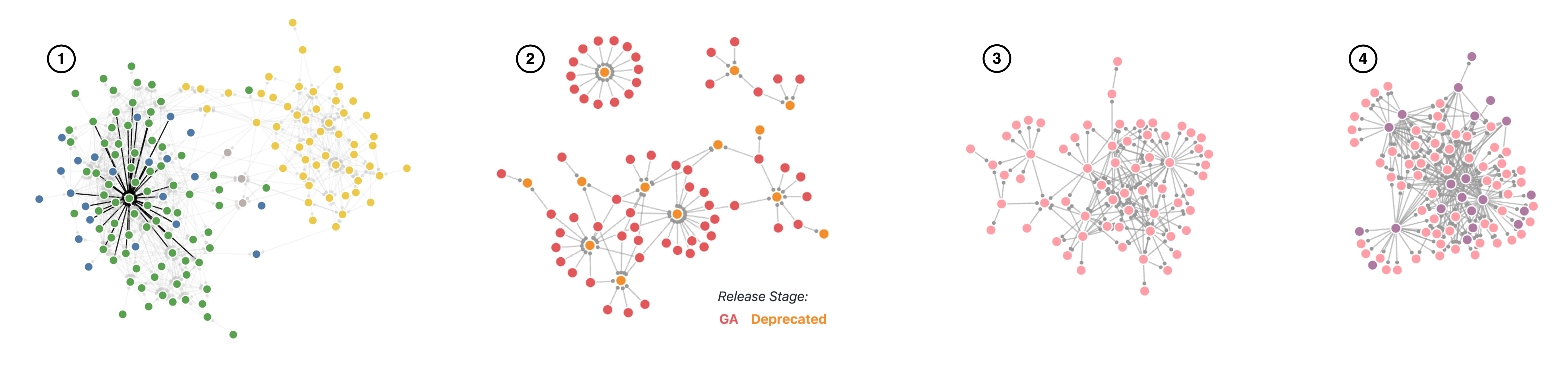}
    \caption{Different aspects that our tool can visualize: (1) highlighting a central logging service in the organization from Figure~\ref{fig:all-services}, (2) GA services directly depending on recently deprecated (but not yet retired) services, (3) dependencies of services in one unit solely having organization-internal dependencies, (4) dependencies of a unit relying also on services provided by other organizations.}
    \label{fig:collection}
\end{figure*}

It is worth noting that there were some individual green services that were more closely tied to yellow (legacy) services than to other green services.
We investigated those green services and noticed that, while not belonging to the legacy environment, those services had already been marked as deprecated and will hence be migrated soon.
We further noticed a strong relationship between service dependencies and organizational structure, a phenomenon we investigate more closely in the following section (use case \textsc{Organization}).
A feature that proved helpful in exploring individual services and their role in the overall service landscape was the edge highlighting feature that we implemented (see Figure~\ref{fig:collection}-1): while a left-click on a service opens the metadata panel, a right-click highlights all incoming service dependencies, i.e., all services depending on the selected service.

\begin{roundedbox}
\noindent\textsc{Consumption}: The force-directed layout proved useful in revealing a strong coupling between sets of services.
\end{roundedbox}

Managing internal service metadata in a large international software company is a complex task. As mentioned above, such metadata usually originates from different source systems. This, together with the constant evolution of the services, leads to potential data \textsc{Quality} issues.
Our visualization proved useful in detecting different kinds of such issues.
Services without dependencies were immediately visible because the force-directed layout arranged them in an outer ring around the connected services in the center.
Furthermore, users were able to detect missing data using the gray color that we assigned to nodes with missing values.
In the cases we analyzed, missing data could usually be attributed to the fact that services were already deprecated, retired, or peripheral to the cloud platform.

\begin{roundedbox}
\noindent\textsc{Quality}: The force-directed layout and the node color coding proved useful for detecting data quality issues.
\end{roundedbox}

\section{Extended Visualization Tool}
\label{sec:visualization-extended}

After our initial visualization tool was available, colleagues from a different team approached us with a number of questions about SAP's cloud services.
While we cannot share the detailed questions, we will outline aspects of the service metadata they were interested in and how we adapted our tool to support the additional use cases.

\subsection{Design Decisions}

Our colleagues had much more detailed questions about service \textsc{Consumption} within subsets of services, filtered by cloud environment, organizational unit, or release stage.
They were particularly interested in service \textsc{Deprecation}, because deprecated services with few or no dependencies are candidates for service retirement.
Another important aspect was the \textsc{Organizations} responsible for the services and downstream \textsc{Reporting} tasks.
To support these new use cases, we extended our tool.
Previously, it was possible to filter services, e.g., by their cloud environment or organizational unit.
However, it was not possible to independently filter both sides of the dependency relationship.
For example, it was not possible to select all services from a particular organizational unit that depended on a deprecated service.
Therefore, we significantly extended the filter options and added a graphical user interface in addition to the URL parameters.
We continued to encode the filter options in the URL to enable sharing of filtered views.
For convenience, we added an autocompletion feature to the service ID and service name filters.

To make it clear which services were matched by the service filter and which were dependencies of those filtered services, we optionally highlighted the matched services using a thick black circle (see, e.g., Figure~\ref{fig:service-815-viz}-2).
The non-marked services are the direct and transitive dependencies of the marked filtered services. 
We further added an option to hide services without any service dependencies to reduce clutter.
The possibility to show dependencies only up to a certain depth existed before and proved useful, because some of our colleagues' questions required focusing, e.g., on direct dependencies only (corresponding to a max depth of 1, see Figure~\ref{fig:service-815-viz}-1).
Moreover, we added an option to encode the service release stage in the node color (see Figure~\ref{fig:collection}-2) and added the possibility to invert the dependency direction, i.e., moving from the default \emph{requires} relation to a \emph{required by} relation (see Figures~\ref{fig:service-815-viz}-3).
To support the \textsc{Reporting} use case, we added an export feature that created two CSV files: one with the metadata of the currently filtered services and one with their dependencies.

To help users understand the evolution of a service as part of the \textsc{Deprecation} use case, we needed to retrieve the time stamps of release stage changes from a different system and integrate this information with the existing service metadata.
We integrated the resulting release stage change log in the service metadata panel (see panel in Figure~\ref{fig:service-815-viz}).

\subsection{Observations}

Figure~\ref{fig:collection}-2 shows \emph{GA} services (red nodes) directly depending on services in release stage \emph{Deprecated} (orange nodes).
This reveals that there are indeed \emph{GA} services directly depending on recently deprecated---but not yet retired---services.
As mentioned before, service deprecation is a logical first step that triggers service migration before the deprecated service can be retired.
The release stage change log enabled our users to explore this temporal relationship.

\begin{roundedbox}
\noindent\textsc{Deprecation}: Selecting only direct dependencies and adding temporal information enabled users to identify deprecated services that are candidates for retirement.
\end{roundedbox}

Our colleagues were particularly interested in the organizational perspective on service dependencies.
While we cannot go into the details of which organizations they were focusing on, we can report some general observations.
Figures~\ref{fig:collection}-3 and \ref{fig:collection}-4, for instance, show the dependencies of two organizational units side by side.
The node color encodes the high-level organization they are part of.
The first unit (Figure~\ref{fig:collection}-3) has mainly organization-internal dependencies.
Services provided by the second organization (Figure~\ref{fig:collection}-4), on the other hand, were intertwined with services provided by another high-level organization (purple).
Often, such patterns corresponded to units providing foundational services versus units maintaining business services on top.
We see this as an indication of Conway's law, which states that ``any organization that designs a system (defined broadly) will produce a design whose structure is a copy of the organization's communication structure''~\cite{Conway1968}.
Such patterns are not problematic as long as they are---as in the above examples---a function of the organization's purpose and not a mere artifact of the hierarchical structure.

\begin{roundedbox}
\noindent\textsc{Organization}: Certain service dependency patterns were explainable by the organizational structure.
\end{roundedbox}

Regarding the \textsc{Reporting} use case, it was important to understand the work mode of our users and what their deliverables were.
Our goal was to provide value beyond the capabilities of the existing tooling without replicating existing functionality.
Hence, we decided to implement only very basic reporting features (e.g., number of matched services/dependencies), leaving downstream analysis tasks to the tools our users were accustomed to.
The CSV export that we added provided a clean interface between our tool, which was used to explore, filter, and visualize the data, and the analytics software used for further analysis and reporting.

\section{Discussion and Conclusion}

In this paper, we reported findings from an industrial case study on managing service dependencies using a visualization tool.
The highlighted use cases demonstrate that a tailored yet relatively simple graph visualization approach can have substantial impact on understanding and maintaining a company's software service landscape.
Our visualization provides an overview of hundreds of services by default, but interactivity and adaptability were key for gaining deeper insights.
In a corporate setting, it is important to recognize that a novel (visualization) tool is usually part of a larger analysis and reporting process.
In our case, this meant consuming and preprocessing data from existing systems and providing an interface for users to export data for further analysis.
We incrementally increased the value of our visualization tool, starting with a quickly built prototype that already provided useful results, serving as a minimum viable solution.
This enabled us to collect user feedback early on, resulting in additional and more specific feature requests.
Although this process led to a tool tailored to the service metadata that SAP collects and the use cases of SAP-internal stakeholders, the process itself and our general visualization approach do likely generalize to other companies.
In addition, our use cases can act as candidate scenarios for similar applications.

\bibliographystyle{IEEEtran}
\bibliography{literature}

\begin{thebibliography}{10}
\providecommand{\url}[1]{#1}
\csname url@samestyle\endcsname
\providecommand{\newblock}{\relax}
\providecommand{\bibinfo}[2]{#2}
\providecommand{\BIBentrySTDinterwordspacing}{\spaceskip=0pt\relax}
\providecommand{\BIBentryALTinterwordstretchfactor}{4}
\providecommand{\BIBentryALTinterwordspacing}{\spaceskip=\fontdimen2\font plus
\BIBentryALTinterwordstretchfactor\fontdimen3\font minus
  \fontdimen4\font\relax}
\providecommand{\BIBforeignlanguage}[2]{{%
\expandafter\ifx\csname l@#1\endcsname\relax
\typeout{** WARNING: IEEEtran.bst: No hyphenation pattern has been}%
\typeout{** loaded for the language `#1'. Using the pattern for}%
\typeout{** the default language instead.}%
\else
\language=\csname l@#1\endcsname
\fi
#2}}
\providecommand{\BIBdecl}{\relax}
\BIBdecl

\bibitem{oasis-soa-rm2006}
C.~M. MacKenzie, K.~Laskey, F.~McCabe, P.~F. Brown, and R.~Metz, ``{OASIS
  Reference Model for Service Oriented Architecture 1.0},''
  http://docs.oasis-open.org/soa-rm/v1.0/soa-rm.html, July 2006.

\bibitem{Vogels2022}
W.~Vogels, ``{The Distributed Computing Manifesto},''
  https://www.allthingsdistributed.com/2022/11/amazon-1998-distributed-computing-manifesto.html,
  November 2022.

\bibitem{DBLP:conf/icsm/FritzschB0Z19}
J.~Fritzsch, J.~Bogner, S.~Wagner, and A.~Zimmermann, ``Microservices migration
  in industry: Intentions, strategies, and challenges,'' in \emph{{ICSME}
  2019}.\hskip 1em plus 0.5em minus 0.4em\relax {IEEE}, pp. 481--490.

\bibitem{DBLP:conf/softvis/Deursen10}
A.~van Deursen, ``A pragmatic perspective on software visualization,'' in
  \emph{{SOFTVIS} 2010}.\hskip 1em plus 0.5em minus 0.4em\relax {ACM}, 2010,
  pp. 1--2.

\bibitem{bavota2013evolution}
G.~Bavota, G.~Canfora, M.~Di~Penta, R.~Oliveto, and S.~Panichella, ``The
  evolution of project inter-dependencies in a software ecosystem: The case of
  {Apache},'' in \emph{{ICSM} 2013}.\hskip 1em plus 0.5em minus 0.4em\relax
  IEEE, pp. 280--289.

\bibitem{kikas2017structure}
R.~Kikas, G.~Gousios, M.~Dumas, and D.~Pfahl, ``Structure and evolution of
  package dependency networks,'' in \emph{{MSR} 2017}.\hskip 1em plus 0.5em
  minus 0.4em\relax IEEE, pp. 102--112.

\bibitem{kula2018developers}
R.~G. Kula, D.~M. German, A.~Ouni, T.~Ishio, and K.~Inoue, ``Do developers
  update their library dependencies? {A}n empirical study on the impact of sec.
  advisories on library migration,'' \emph{EMSE}, vol.~23, 2018.

\bibitem{DBLP:journals/queue/EsparrachiariRR18}
S.~Esparrachiari, T.~Reilly, and A.~Rentz, ``Tracking and controlling
  microservice dependencies,'' \emph{{ACM} Queue}, vol.~16, no.~4, p.~10, 2018.

\bibitem{10.1016/j.jss.2014.03.071}
M.~Shahin, P.~Liang, and M.~A. Babar, ``A systematic review of software
  architecture visualization techniques,'' \emph{JSS}, vol.~94, 2014.

\bibitem{DBLP:phd/de/Beck2013}
F.~Beck, ``Understanding multi-dimensional code couplings,'' Ph.D.
  dissertation, University of Trier, Germany, 2013.

\bibitem{10.1109/vissof.2011.6069454}
U.~Erdemir, U.~Tekin, and F.~Buzluca, ``{E-Quality}: a graph based object
  oriented software quality visualization tool,'' in \emph{{VISSOFT} 2011}.

\bibitem{10.1145/1409720.1409735}
J.~Dietrich, V.~Yakovlev, C.~McCartin, G.~Jenson, and M.~Duchrow, ``Cluster
  analysis of {Java} dependency graphs,'' in \emph{{SoftVis} 2008}.

\bibitem{zhou2018fault}
X.~Zhou, X.~Peng, T.~Xie, J.~Sun, C.~Ji, W.~Li, and D.~Ding, ``Fault analysis
  and debugging of microservice systems: Industrial survey, benchmark system,
  and empirical study,'' \emph{IEEE TSE 2018}, vol.~47, no.~2.

\bibitem{DBLP:conf/sigsoft/Guo0WLJDXS20}
X.~Guo, X.~Peng, H.~Wang, W.~Li, H.~Jiang, D.~Ding, T.~Xie, and L.~Su,
  ``Graph-based trace analysis for microservice architecture understanding and
  problem diagnosis,'' in \emph{{ESEC/FSE} 2020}.\hskip 1em plus 0.5em minus
  0.4em\relax {ACM}, pp. 1387--1397.

\bibitem{DBLP:conf/bir/GaidelsK20}
E.~Gaidels and M.~Kirikova, ``Service dependency graph analysis in microservice
  architecture,'' in \emph{{BIR} 2020}, vol. 398, pp. 128--139.

\bibitem{10.1109/ficloud57274.2022.00042}
H.~Farsi, D.~Allaki, A.~En-Nouaary, and M.~Dahchour, ``A graph-based solution
  to deal with cyclic dependencies in microservices architecture,'' in
  \emph{{FICLOUD} 2022}.

\bibitem{10.1109/sose55356.2022.00011}
T.~Cerny, A.~S. Abdelfattah, V.~Bushong, A.~Al~Maruf, and D.~Taibi,
  ``Microservice architecture reconstruction and visualization techniques: a
  review,'' in \emph{{SOSE} 2022}.

\bibitem{Conway1968}
M.~E. Conway, ``How do committees invent,'' \emph{Datamation}, April 1968.

\end{thebibliography}
\end{document}